# Mesoscopic Elastic Distortions in GaAs Quantum Dot Heterostructures

*Anastasios Pateras[1], Joonkyu Park[1], Youngjun Ahn[1], Jack A. Tilka[1], Martin V. Holt[2], Christian Reichl[3], Werner Wegscheider[3], Timothy A. Baart[4], Juan Pablo Dehollain[4], Uditendu Mukhopadhyay[4], Lieven M. K. Vandersypen[4], and Paul G. Evans[1,*]*

[1] Department of Materials Science & Engineering, University of Wisconsin-Madison, Madison, Wisconsin 53706 USA

[2] Center for Nanoscale Materials, Argonne National Laboratory, Argonne, IL 60439, USA

[3] Laboratory for Solid State Physics, ETH Zürich, Zürich CH-8093, Switzerland

[4] QuTech and Kavli Institute of NanoScience, Delft University of Technology, PO Box 5046, 2600 GA, Delft, The Netherlands

ABSTRACT: Quantum devices formed in high-electron-mobility semiconductor heterostructures provide a route through which quantum mechanical effects can be exploited on length scales accessible to lithography and integrated electronics. The electrostatic definition of quantum dots in semiconductor heterostructure devices intrinsically involves the lithographic fabrication of intricate patterns of metallic electrodes. The formation of metal/semiconductor interfaces, growth processes associated with polycrystalline metallic layers, and differential thermal expansion produce elastic distortion in the active areas of quantum devices. Understanding and controlling these distortions presents a significant challenge in quantum device development. We report synchrotron x-ray nanodiffraction measurements combined with dynamical x-ray diffraction modeling that reveal lattice tilts with a depth-averaged value up to 0.04° and strain on the order of $10^{-4}$ in the two-dimensional electron gas (2DEG) in a GaAs/AlGaAs heterostructure. Elastic distortions in GaAs/AlGaAs heterostructures modify the potential energy landscape in the 2DEG due to the generation of a deformation potential and an electric field through the piezoelectric effect. The stress induced by metal electrodes directly impacts the ability to control the positions of the potential minima where quantum dots form and the coupling between neighboring quantum dots.

Semiconductor heterostructures can host devices that permit electronic quantum states to be precisely controlled.[1] Under appropriate conditions, electronic states have long dephasing times and thus form the basis for emerging quantum electronic technologies.[2-4] One realization of such devices consists of states hosted near interfaces in GaAs/AlGaAs heterostructures at which a two-dimensional electron gas (2DEG) is formed.[2] Quantum dots (QDs) can be electrostatically defined at the 2DEG interface of the two materials by voltages applied to lithographically patterned nanoscale gate electrodes. The metallic gates distort the semiconductor crystal due to stresses transmitted through the metal/semiconductor interface.[5, 6] The piezoelectricity and deformation potential of GaAs lead to a coupling of the electrode-induced strain and the electronic energy landscape, resulting in additional spatially inhomogeneous terms in the electronic Hamiltonian.[5, 7] Similar stress-driven effects pertain to two-dimensional hole systems (2DHSs).[8] These strain effects perturb the electronic landscape and in particular alter the position and depth of the potential minima that in turn determine the location of the quantum dots, the number of electrons residing in the dots (for a given set of gate voltages), and the coupling between adjacent quantum dots. In addition to these unintended stress-related effects, there are ways that interface stress can be deliberately employed in the design of quantum devices. For example, stress effects can be useful in defining devices by aligning the gate voltage-induced potential minima with the stress-induced potential minima, so that stress aids quantum dot formation.[9] The stress can also lead to distributed electronic effects arising from large, micron-scale periodic patterns.[8, 10, 11] The characterization and quantification of the nanoscale variation of stresses in GaAs heterostructures, however, has been a significant experimental challenge and has limited the understanding, control, and eventual exploitation of stress-driven effects in quantum electronics.

Residual stress in metal electrodes is a nearly universal phenomenon arising from microstructural effects in polycrystalline thin films. Stress generation mechanisms during electrode thin film deposition include capillary forces on clusters of metal atoms, grain boundary formation, and the preferential incorporation of atoms at grain boundaries.[12-16] The stress induced by the metallic gates can vary significantly as a function of several interdependent parameters, such as the substrate temperature during growth, grain size, overall film thickness, and deposition rate. The stress also depends on the thermal history after deposition due to the motion of atoms and defects during thermal cycling.[17-19] The magnitude of the metal-induced stress is thus difficult to predict and in principle the same electrode pattern can lead to different residual stresses in quantum devices, if the electrodes are formed under different conditions. In the fabrication of quantum devices on GaAs, the gate electrodes are often formed by a Ti adhesion layer deposited on GaAs and subsequently capped with a thicker Au layer. The formation of stacked thin films in this configuration results in complex gate stress distributions. Au and Au/Ti thin film metallizations on GaAs and other semiconducting substrates have stresses of 50 to 100 MPa, depending on the total electrode thickness and the relative thicknesses of Au and Ti, with lower stresses observed at smaller Au thicknesses.[20, 21] The stress in Pt/Ti thin films on GaAs substrates has approximately the same magnitude as in Au/Ti.[22] The mismatch of the coefficients of thermal expansion of the metal gate electrodes and semiconductor substrate is a further independent source of mechanical deformation through which stress is introduced during cooling the devices to cryogenic temperatures.

Theoretical models for the effect of elastic distortions on the electronic states of the 2DEG describe the variation of the energy of the conduction band edge as a function of the applied stress. The conduction band energy depends on the hydrostatic component of the distortion via the deformation potential[23] and on long-range distortions via the piezoelectric effect.[24, 25] Studies combining electronic transport measurements and theory have evaluated several elastic and electronic models, but comparison with experiment remains imprecise because the effects of strain on the potential landscape are convoluted with the (often larger) effect of background charged impurities. Examples of successes in the study of stresses using transport measurements include determining the relative contributions of several spatial harmonic components of the elastic distortion from magnetoresistance measurements.[5, 25] The magnitude of commensurability oscillations in 2DHSs indicates that the total stress-induced potential varies from approximately 1 to 10% of the hole Fermi energy.[8] Uncertainty about the magnitude and spatial variation of the elastically induced distortions and the degree of screening of the 2DEG, however, complicates the comparison of electronic models with transport studies.[8, 25] In addition, the crystallographic symmetry of GaAs makes modeling the effects of distortions challenging. The influence of the lattice distortions on the piezoelectric potential developed in GaAs/AlGaAs 2DEGs, for example, depends on the crystallographic direction.[25, 26] The piezoelectric effect is negligible for stresses along the [010] direction and maximum along the [110] direction, thus, the contribution to the potential depends on the orientation of the gates.

For geometrically simple patterns, such as linear electrodes, the spatial variation of the nanoscale stress is straightforward and can be accurately predicted using analytical solid mechanics models.[27] Several mechanical models provide accurate predictions of the stress at locations away from the edges of the electrodes. We employ the edge-force mechanical model, which is based on the approximation that the sharp corners of the stressors can be modeled as lines of force.[28] The distortion in the vicinity of an isolated interface stressor has displacements described by a three-dimensional spatially varying strain tensor.[29] More elaborate electrode patterns have strain fields resulting from the superposition of the stresses induced by individual electrodes, making an analytical comparison difficult.

The quantum dot device probed in this work consisted of lithographically patterned gates on an epitaxial GaAs/AlGaAs heterostructure, as shown in Figure 1a. The semiconductor heterostructure is composed of layers grown by molecular beam epitaxy on a (001) GaAs substrate. Starting from the surface, the layers are: a 5 nm-thick GaAs cap, 45 nm of $Al_xGa_{1-x}As$ with x=0.31, a very thin Si-doped layer, 40 nm of AlGaAs with x=0.32, 500 nm of GaAs, a superlattice of 80 periods with a repeating unit of 7 nm AlGaAs/3 nm GaAs, and a 400 nm-thick GaAs buffer layer. The 2DEG forms at the interface between the 40 nm-thick AlGaAs layer and the 500 nm-thick GaAs layer. The tilt of the crystal lattice within the AlGaAs layers is described using the angular definitions shown in Figure 1b. The x-ray studies determine two components of the misorientation of the AlGaAs layers, described by the angles $\phi_{para}$ and $\phi_{perp}$, with respect to the sample surface normal. The gates are composed of a 20 nm-thick Au thin film on a 5 nm-thick Ti adhesion layer and form a complex geometrical pattern defining the quantum device, as illustrated in Figure 1c. A detailed description of the gate fabrication process is given in the experimental section.

The structural investigation of the QDs in the GaAs/AlGaAs heterostructure employed synchrotron x-ray nanobeam diffraction (see Figure 1b), an approach which allows the non-destructive determination of the electrode-induced lattice distortions. The x-ray measurements were conducted at ambient temperature, approximately 290 K. The zone plate x-ray focusing optic introduces a beam divergence of 0.34°, far larger than the divergence of conventional high-resolution laboratory x-ray sources. The reflection of the convergent x-ray beam from the thin film heterostructure results in complex diffraction patterns. Significant progress in interpreting nanobeam diffraction patterns has been made for systems in which the component layers produce signals that can be readily distinguished from the substrate using kinematical scattering theory and phase-retrieval techniques.[30-34] The approximation underpinning the kinematical approach, however, does not apply in GaAs/AlGaAs heterostructures for two reasons. First, these samples consist of crystals with thicknesses larger than the x-ray extinction depth, making multiple scattering important. Second, the GaAs/AlGaAs heterostructures incorporate thin and thick layers that are nearly lattice matched, resulting in complex interference between layers. The lattice mismatch between $Al_xGa_{1-x}As$ and GaAs is $4.16 \times 10^{-4}$ for x=0.3, corresponding to an $Al_xGa_{1-x}As$ lattice constant of $\alpha_2$=5.6556 Å, compared to the GaAs lattice parameter, $\alpha_1$=5.6533 Å.[35] The dynamical diffraction description incorporates multiple scattering effects and provides quantitative insight into the distribution of scattered intensity.

The x-ray diffraction patterns were interpreted using a simulation method in which the heterostructure is modeled as a stack of crystalline layers with different lattice parameters and scattering factors.[36] The simulation combines a nanobeam diffraction model with the dynamical theory of x-ray diffraction to describe the interaction of the focused x-ray nanobeam with the sample and predict the far-field intensity distribution on the detector.[37, 38] Dynamical diffraction considerations are particularly important in the GaAs/AlGaAs system because of the very close lattice match between the GaAs substrate and the AlGaAs layers in the heterostructures.

Figure 1d shows a diffraction pattern acquired with the x-ray beam illuminating an area far from the metal gates. At the nominal incident angle of $\theta_B = 24.95°$ used in Figure 1d, the angular center of the focused beam meets the Bragg condition for the GaAs 004 reflection at 10.4 keV (corresponding to a wavelength $\lambda$=1.19 Å). The key dynamical diffraction features of the diffraction pattern in Figure 1d are the two vertical bright lines of diffracted intensity at the center of the image and which cannot be reproduced using the kinematical theory of x-ray diffraction.[39, 40] The sharp line at a slightly higher value of the angle 2θ, arises from the thick GaAs layers and the substrate. The second sharp line of intensity at lower 2θ arises from the 800 nm-thick GaAs/AlGaAs superlattice, which has a lattice parameter $d_{avg}$=5.6545 Å. Here 2θ is defined for each position on the detector as the angle between the central ray of the focused beam and the ray of the x-rays scattered to that position. The lattice constant was calculated by fitting the positions of the peak lines from the superlattice and substrate using the dynamical diffraction simulation. The angular divergence of the focused x-ray beam leads to the formation of a disk of intensity modulated by the angular dependence of the x-ray reflectivity of the heterostructure. The dark circular feature at the center of Figure 1d is the shadow of the center stop. The features observed in the experimental diffraction patterns that originate from the heterostructure are reproduced in the simulated diffraction pattern shown in Figure 1e. The concentric dark rings in the experimental diffraction pattern arise from fabrication artifacts in the zone plate.[41]

Figure 2a shows a second diffraction pattern acquired at a nominal incident angle of 25.2°, acquired with the sample slightly rotated away from the nominal Bragg condition for the GaAs 004 reflection. This angle reduces the intensity of the reflection from the thick GaAs substrate because its Bragg condition falls outside the primary cone of the focused x-ray radiation. The AlGaAs layers above the 2DEG produce a series of intensity fringes in each diffraction pattern due to the interference of x-rays incident at different angles in the focused beam. The simulated diffraction pattern for the same angle of incidence is plotted in Figure 2b. The two bright lines originating from the thick GaAs layers and the substrate can be seen outside the nominal cone of the diffracted intensity. These are observed in the experiment because the focused beam has angular tails extending outside this cone. The features in Figure 2b can also be interpreted using the intensity distribution predicted by the Darwin theory of dynamical diffraction for a plane wave of incident x-rays. The predicted intensity is for a plane-wave incident beam is plotted in Figure 2c as a function of the x-ray scattering wavevector $Q_z=(4\pi/\lambda) \sin\theta$, where $\theta$ is the plane-wave incident angle. The sharp region of total reflection for the 004 GaAs Bragg peak of a semi-infinite substrate, inset in Figure 2c, is the feature giving rise to one of the bright, sharp reflections observed in the experimental diffraction patterns.

The lattice tilts along the beam footprint direction can be measured by comparing the measured and simulated diffraction patterns and tracking the angular displacement of these intensity fringes. The tilt of the AlGaAs layer was measured by tracking the thickness fringe between $2\theta=50.19°$ and $50.24°$, indicated with an arrow in Figure 2a. In the orthogonal direction of the diffraction pattern, a shift by an angle $\Delta 2\theta$ along the $2\theta$ direction which is parallel to the footprint corresponds to a tilt of $\phi_{para} = ½\ \Delta 2\theta$. A shift of the detected intensity along the vertical axis of the diffraction patterns by an angle $\Delta\chi$, corresponds to a tilt of the lattice planes $\phi_{perp} = \Delta\chi/(2\sin\theta B)\approx 1.19\Delta\chi$. Under the condition that all of the (004) atomic planes in the layer contribute in phase to the x-ray diffraction (i.e. near the peak intensity of the diffraction curve) the centroid angular displacement along the $\chi$ direction, $\Delta\chi_{center}$, is given by $\Delta\chi_{center} = (1/1.19)\ \phi_{perp, mean}$.

In order to understand the nanoscale variation of the distortion induced by the electrodes, we first consider the distortion induced by a linear Au/Ti electrode. The symmetry of the electrode constrains the direction of lattice tilts to be such that the shift of the diffracted intensity is primarily along the $\chi$ direction of the diffraction pattern.[42] The shift was determined by summing the intensity of the thickness fringe inside the rectangular area shown in Figure 2a along 2θ and tracking the maximum of the distribution along χ. Figure 3a shows a map of the tilts along the χ direction as a function of the beam position across the electrode.

The spatial distribution of the distortion resulting from an isolated rectangular electrode was compared to an analytical prediction based on the edge-force model.[27] To simplify the calculation we have used the analytical result presented in ref. 28 and approximated the GaAs layer as an isotropic elastic solid with Young's modulus E=85.5 GPa and Poisson's ratio ν=0.31.[35] The tilts predicted by the calculation were averaged over the thickness of the AlGaAs layer, from the surface to a depth of 90 nm, and convoluted with a Gaussian resolution function with a full-width at half maximum (FWHM) of 82 nm. This resolution function represents an increase in the effective size of the focused x-ray beam on the sample due to a slight (but unknown) displacement of the sample from the focus. The residual stress in the model was adjusted so that the magnitudes of the tilts match the experimental values. The value of the residual stress that provided the best fit for the tilt distribution in Figure 3b was 57 MPa. The fit shown in Figure 3b also accounts for the long-length scale curvature of the AlGaAs layer due to curvature of the GaAs substrate. Such average tilts can arise from an overall curvature of the substrate due to stresses imposed during high-temperature growth.

The residual stress in the electrodes imparts a tensile stress on the AlGaAs/GaAs heterostructure. Models considering only differential contraction effects due to cooling to cryogenic temperatures predict compressive stresses due to the larger coefficient of thermal expansion of Au in comparison with GaAs.[5, 25] The thermal contraction effect has a smaller magnitude and opposite sign in comparison with the residual stress reported here. A calculation based on the mismatch of the coefficients of thermal expansion of GaAs and Au, using the approach in ref. 25, predicts that the strain is reduced by approximately a factor of two at cryogenic temperatures. Previously reported electronic transport studies are consistent with a total tensile stress, which is what would occur through the sum of the effects of the residual stress and the contribution from differential contraction.[25, 43]

The results of the mechanical model also provide the magnitude of the in-plane strain at the depth where the 2DEG forms. The computed strain at this depth is shown in Figure 3c, where for a 100 nm-wide Au/Ti gate, the in-plane strain ($\varepsilon_{xx}$) at a depth of 90 nm below the surface of the heterostructure is plotted as a function of the lateral distance from the gate center. The induced in-plane strain is tensile underneath the gate, with a maximum magnitude of $4 \times 10^{-5}$, and changes sign within 200 nm from the gate center.

The QD region incorporates a more intricate gate pattern consisting of multiple metal electrodes. Due to the asymmetry of the geometrical pattern of the gates, the directions and magnitude of the electrode-induced distortions are complex. The positions of the electrodes were determined by recording diffraction patterns as a function of the beam position and plotting the integrated intensity within the angular region indicated by the arrow in Figure 2a. A map of this intensity is shown in Figure 4a. The integrated intensity is larger near the gates because of the increased angular acceptance of x-ray reflections in this region due to strain and tilts in the AlGaAs layers.

The largest magnitude tilts along $\chi$ are found at the edges of the metallic gates and in regions where there are several gates closely spaced. A spatial map of the lattice tilts along $\chi$ in the region of the QD, determined using the angular shift of the AlGaAs intensity fringe indicated in Figure 2a, is shown in Figure 4b. The long-lengthscale curvature of the GaAs and superlattice layers was measured by extracting the shift of the intensity distribution of the superlattice peak along $\chi$ and subtracted from the tilts. The two bright lines do not shift along the $2\theta$ direction, indicating that the substrate curvature in this area falls primarily along the direction spanned by $\chi$. We have measured the tilts in the quantum dot region in the rotation described by $\phi_{perp}$ and found tilts ranging from -0.04° to +0.04°. The maximum values of the tilt angles for both angular tilt directions across the quantum dot are of similar magnitudes.

From the diffraction measurements, we have observed stresses leading to an in-plane strain of $4 \times 10^{-5}$ at the depth of the 2DEG beneath isolated electrodes. The magnitudes are similar to those of strains previously intentionally introduced into GaAs quantum wells using surface features in studies of commensurability oscillations and the fractional quantum Hall effect.[8, 26] The strain in the vicinity of the QDs is more difficult to estimate because of the higher geometric complexity of the electrodes, but is likely to be of a similar magnitude to the strain near the isolated electrodes.

Strains of the magnitude observed at the depth of the 2DEG have the potential to be an important factor in quantum device design through two physical mechanisms. The first is the piezoelectric effect, in which the stress in the GaAs and $Al_xGa_{1-x}As$ layers induces a piezoelectric potential.[24, 44] For an average strain on the order of $4 \times 10^{-5}$ along the [110] crystallographic direction we obtain a piezoelectric potential offset between the gate and quantum dots of $\Delta V_{PE} = 5$ mV (see Supporting Information). The piezoelectric potential effectively adds to the voltages on the gate electrodes by an amount that depends on the strain and can therefore lead to uncertainty in the location and shape of the electrostatically defined quantum dots.[45] The second mechanism is the deformation potential that develops when the crystal periodicity changes as in the case of a strained lattice, leading to an energy shift $\Delta E_c^\Gamma = -0.04$ meV.[24]

In addition to the stress arising from the metallization, similar stresses may arise from features associated with the growth of the heterostructure. GaAs heterostructures can include growth-related features that have an impact on electronic states.[46] The dynamical x-ray diffraction approach described here allows nanoscale distortions to be discovered and quantified in lattice-matched thin films such as GaAs/AlGaAs heterostructures. Ultimately, the development of techniques that promote the quantitative understanding of interface stress effects will be crucial for the development and optimization of quantum electronic devices.

**Coherent x-ray Bragg nanodiffraction measurements:** Nanodiffraction studies were conducted using the hard x-ray nanoprobe at station 26-ID of the Advanced Photon Source at Argonne National Laboratory. An x-ray beam with a photon energy of 10.4 keV was selected by a two-bounce Si 111 monochromator and focused to a spot size with a nominal 30 nm FWHM. The focusing optics consisted of a 150 μm-diameter Fresnel zone plate with a 20 nm outermost zone width. The sample was placed at the first order of focus of the zone plate, the direct beam was blocked by a 60 μm-diameter center stop, and the radiation focused to other orders was eliminated by an order-sorting aperture. The nominal incidence angle of the x-ray nanobeam was approximately 25°, close to the Bragg condition for the GaAs 004 x-reflection. Here we have defined a nominal incident angle with respect to the direction of the center of the convergent focused beam. The diffracted beam intensity was recorded using a pixel-array detector (Pixirad-1, PIXIRAD Imaging Counters s.r.l.) placed at distance R=1.012 m from the x-ray focal spot. The detector recorded the x-ray intensity in a plane spanned by angles $2\theta$ and $\chi$, $2\theta$ is along the conventional diffraction angle in the diffraction plane, and $\chi$ is normal to the beam footprint direction.

**Gate deposition:** The semiconductor surface was prepared using an oxygen plasma (10 s),

rinsed in acetone (2 min) and isopropyl alcohol (IPA) (30 s), and dried using $N_2$ gas. The resist was 2 wt. % polymethylmethyacrylate (molecular weight of 950 K) in anisole, which was spin-coated at 2000 rpm (55 s) to form a layer with a thickness of 90 nm, and baked at 175 ºC (10 min). The patterned area was exposed to 100 kV electron beam with the smallest available spot size (2-3 nm) at a beam current of 400 pA. The dose was selected by performing a dose test around 750 $\mu C/cm^2$ and selecting the dose that resulted in gate dimensions closest to the design. The gates were evaporated at rates of 0.1 nm/s for the Ti adhesion layer and 0.1-0.2 nm/s for Au. The resist lift-off was performed after 3 h in warm acetone (53 °C). The sample was then rinsed in IPA (30 s) and dried with $N_2$ gas.

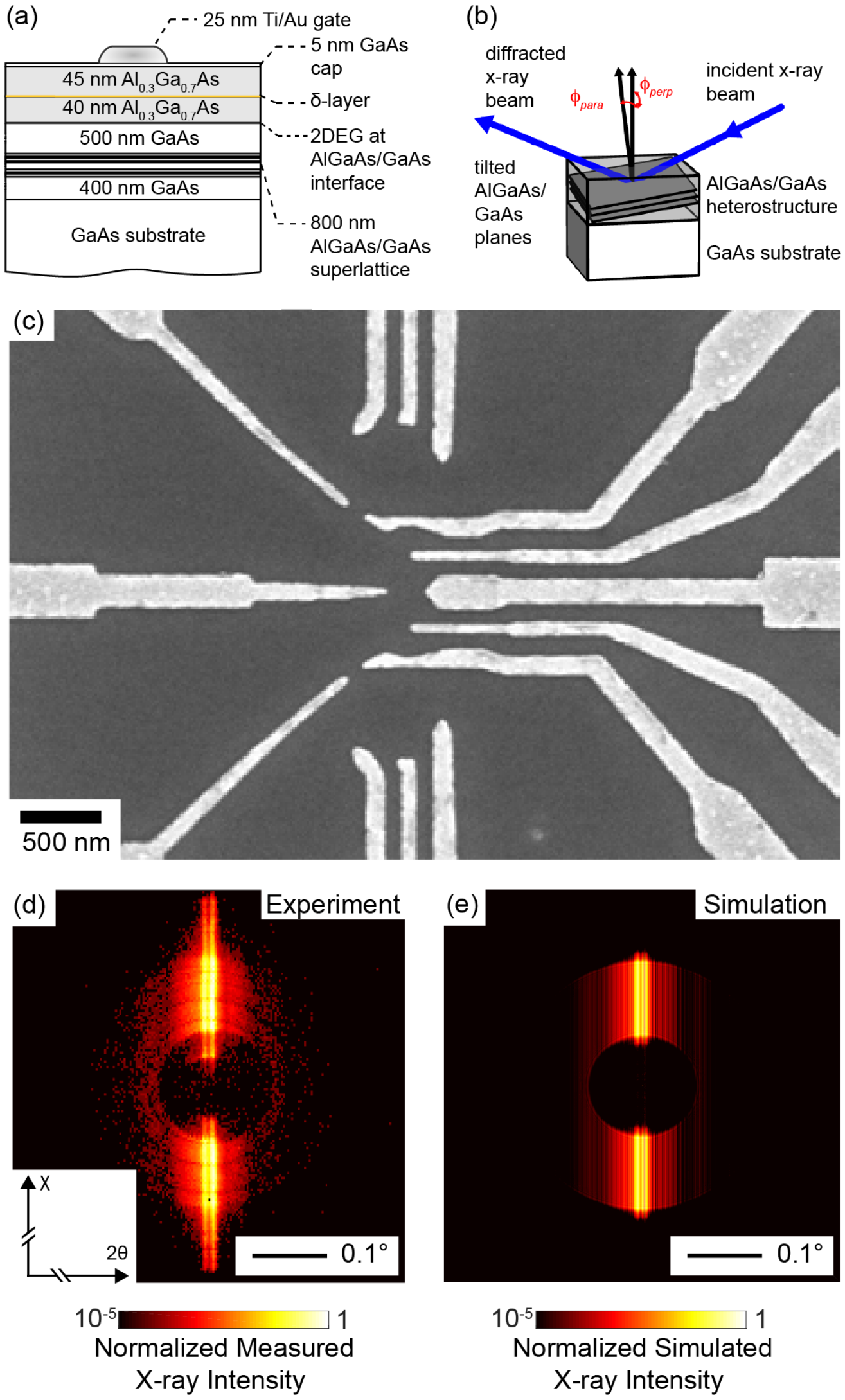

**Figure 1.** (a) Cross-sectional schematic of the GaAs/AlGaAs quantum dot heterostructure. (b) Definitions of angular directions of the stress-induced tilts of crystallographic planes within the AlGaAs layers. (c) Scanning electron micrograph of the electrode pattern in the QD region. (d) Experimental and (e) simulated diffraction patterns of the GaAs/AlGaAs heterostructure acquired at nominal incident angles of 24.95° at which the angular center of the focused x-ray beam meets the Bragg condition for the GaAs 004 reflection.

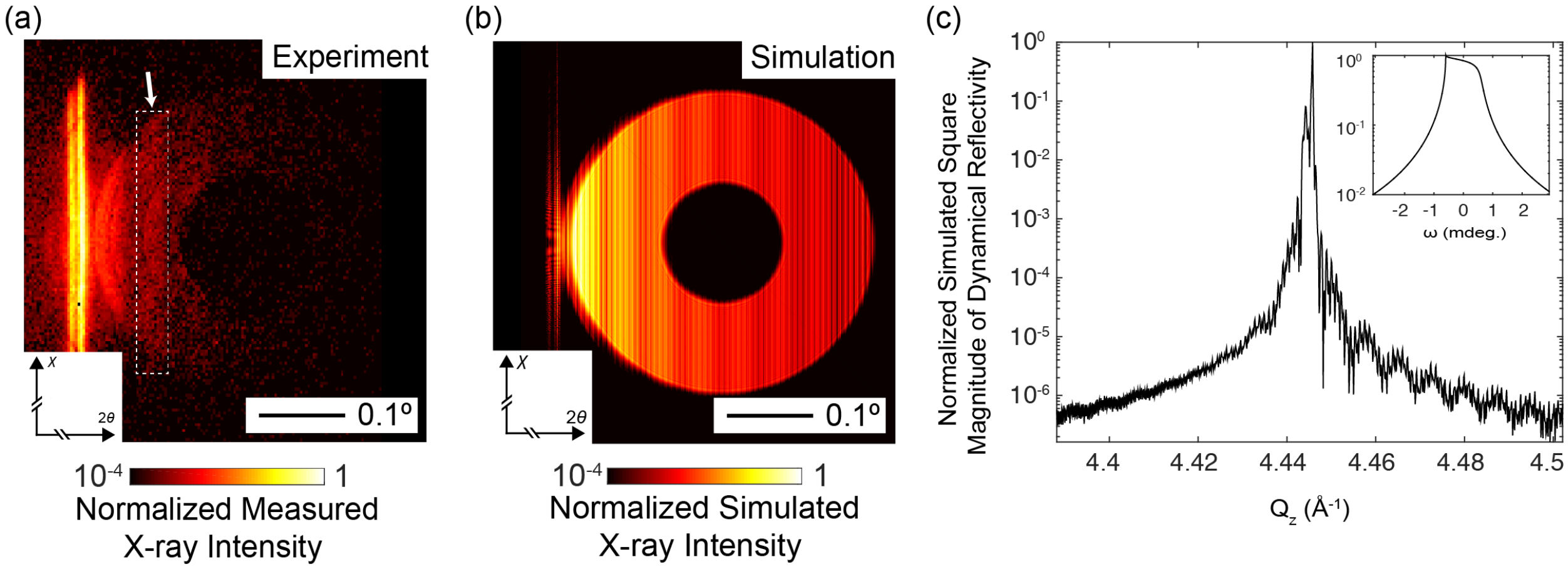

**Figure 2.** (a) Diffraction pattern acquired at a location far from the Au/Ti surface gates with a nominal x-ray incident angle of 25.2°. The arrow indicates a fringe at values of 2θ between 50.26° and 50.31° discussed in detail in the following figures. (b) Simulated diffraction pattern from an untilted heterostructure with the layers depicted in Figure 1a. (c) Predicted normalized diffracted intensity of the quantum dot heterostructure as a function of wavevector $Q_z$ in the case of a plane wave incident x-ray beam. The inset shows the profile of the Darwin dynamical reflectivity for the 004 Bragg reflection of the substrate as a function of ω, the difference between the x-ray plane wave incidence angle and the central angle of the intensity maximum. The sharp features in the reflectivity near $Q_z$=4.445 $Å^{-1}$ produce the sharp features experimentally observed in the scattered x-ray intensity in Figures 1d and 2a.

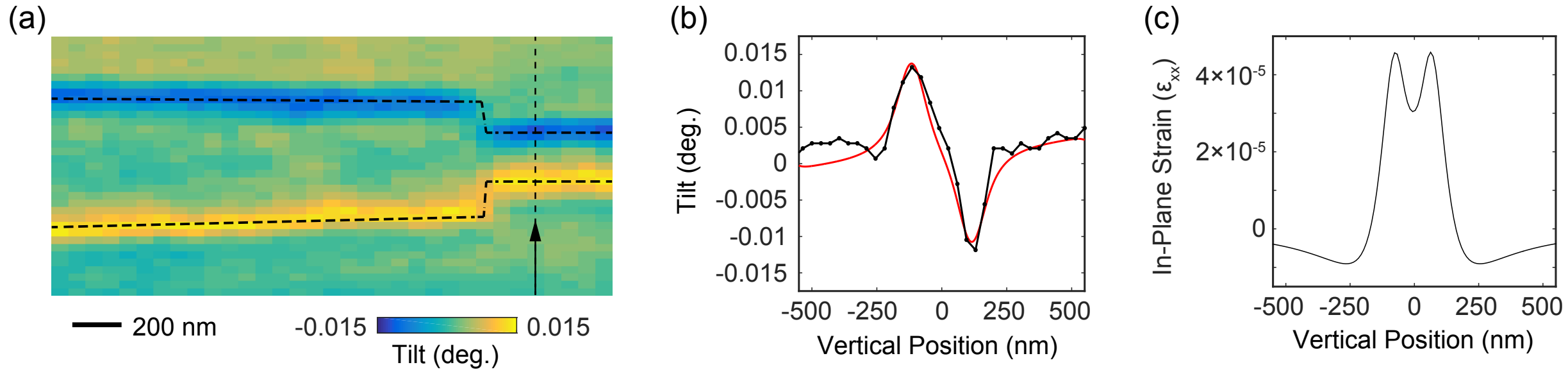

**Figure 3.** (a) Map of the lattice tilts along the normal to the beam footprint direction χ for a linear electrode. The black dashed lines indicate the electrode contours. (b) Tilts measured across the linear electrode following the dashed line indicated on (a). The fit of the depth-averaged, resolution broadened edge-force model if shown by the red line. (c) In-plane strain ($\varepsilon_{xx}$) predicted from the edge-force model, 90 nm below the surface of the film, beneath a 100 nm-wide Au/Ti gate with stress 57 MPa, at the depth where the 2DEG is formed.

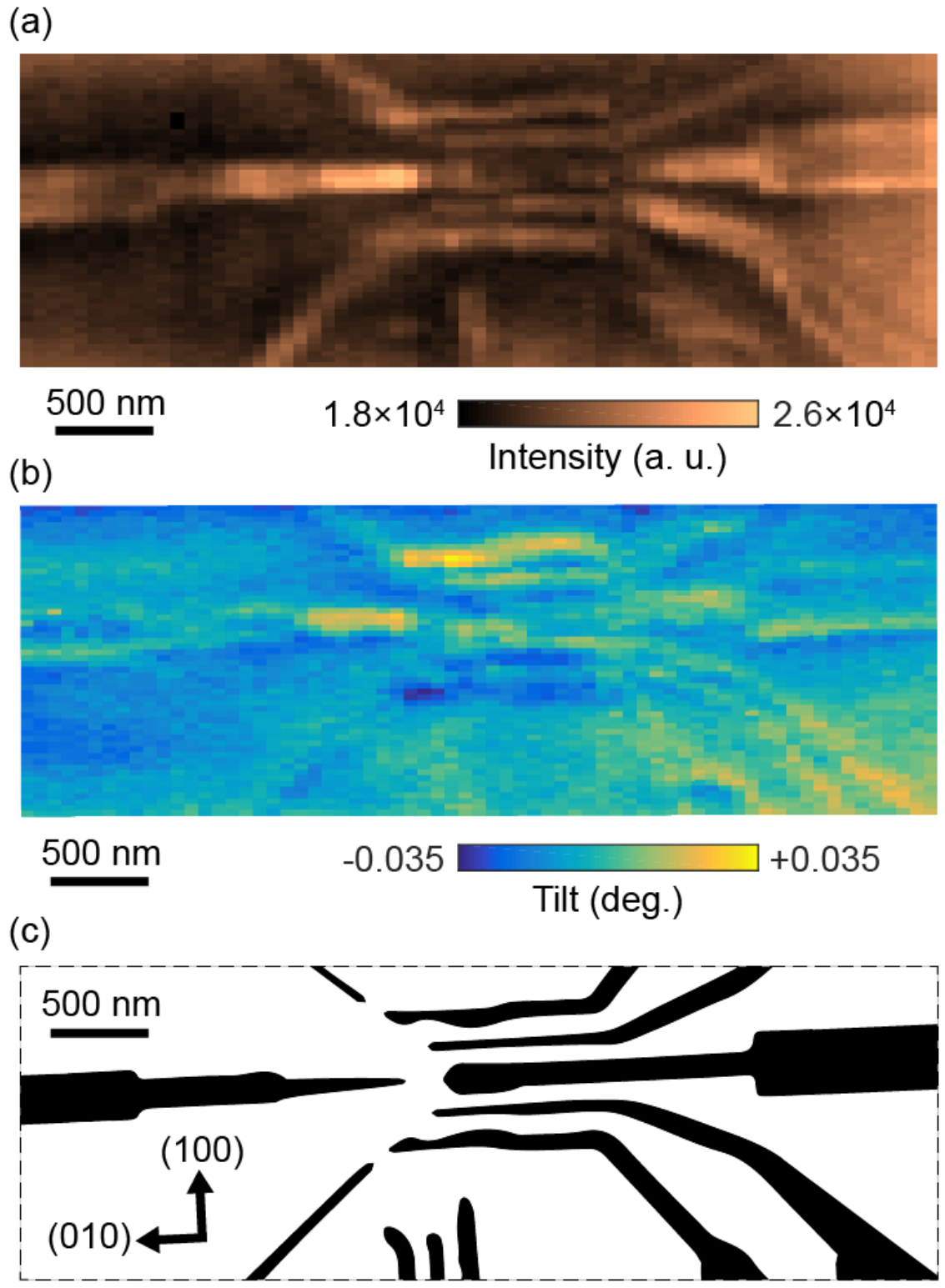

**Figure 4.** (a) Map of the integrated diffracted x-ray intensity in the QD region. Each of the pixels is obtained by integrating the diffracted intensity within the angular range of the intensity fringe appearing in the range of 2θ from 50.26° to 50.31°, inside the dashed box in Figure 2a. (b) Map of the tilts along $\chi$ in the QD region, measured by determining the shift along the $\chi$ direction of the thickness fringe indicated by the dashed box in Figure 2a. (c) Diagram with the locations and orientations with respect to the crystal lattice of the electrodes in this region.

ASSOCIATED CONTENT

**Supporting Information**. The following files are available free of charge.

Calculation of the potential difference generated due to gate-induced stress through the piezoelectric effect inside the AlGaAs layers and the shift of the conduction band minima near the Γ point of GaAs due to the strain dependence of the deformation potential in the Hamiltonian. (PDF)

AUTHOR INFORMATION

**Corresponding Author**

* E-mail: pgevans@wisc.edu

**Funding**

A.P., J.P., Y.A., and P.G.E. were supported by the U.S. DOE, Basic Energy Sciences, Materials Sciences and Engineering, under Contract No. DE-FG02-04ER46147 for the x-ray scattering studies and analysis. J.A.T. acknowledges support from the National Science Foundation Graduate Research Fellowship Program under Grant No. DGE-1256259. Use of the Center for Nanoscale Materials and the Advanced Photon Source, both Office of Science user facilities, was supported by the U.S. Department of Energy, Office of Science, Office of Basic Energy Sciences, under Contract No. DE-AC02-06CH11357. Laboratory characterization at the University of Wisconsin-Madison used instrumentation supported by the National Science Foundation through the UW-Madison Materials Research Science and Engineering Center (DMR-1121288 and DMR-1720415). Work at TU Delft was supported by the Netherlands Organization of Scientific Research (NWO).

TABLE OF CONTENTS GRAPHIC

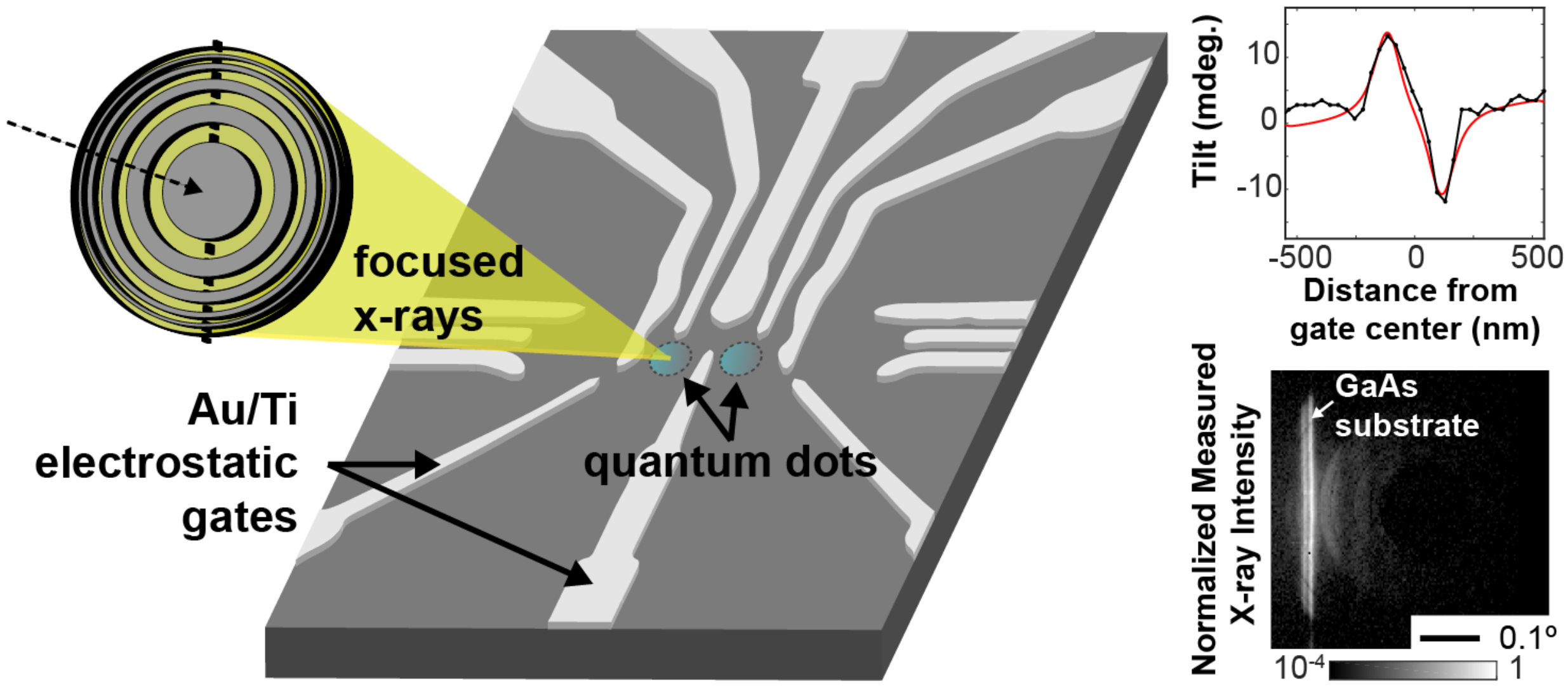

# Supporting Information for:

# Mesoscopic Elastic Distortions in GaAs Quantum Dot Heterostructures

*Anastasios Pateras[1], Joonkyu Park[1], Youngjun Ahn[1], Jack A. Tilka[1], Martin V. Holt[2], Christian Reichl[3], Werner Wegscheider[3], Timothy A. Baart[4], Juan Pablo Dehollain[4], Uditendu Mukhopadhyay[4], Lieven M. K. Vandersypen[4], and Paul G. Evans[1,*]*

[1] Department of Materials Science & Engineering, University of Wisconsin-Madison, Madison, Wisconsin 53706 USA

[2] Center for Nanoscale Materials, Argonne National Laboratory, Argonne, IL 60439, USA Solid

[3] State Physics Laboratory, ETH Zürich, Zürich CH-8093, Switzerland

[4] QuTech and Kavli Institute of NanoScience, Delft University of Technology, PO Box 5046, 2600 GA, Delft, The Netherlands

## 1. Piezoelectric potential in $Al_xGa_{1-x}As$

Ionic crystals that lack a center of symmetry exhibit piezoelectricity.[1,2] For materials with the zinc-blende structure, such as GaAs and AlGaAs, the piezoelectric tensor is:[3]

$$\hat{e}_{ik} = \begin{bmatrix} 0 & 0 & 0 & e_{14} & 0 & 0 \\ 0 & 0 & 0 & 0 & e_{14} & 0 \\ 0 & 0 & 0 & 0 & 0 & e_{14} \end{bmatrix}.$$

Applied stresses result in an electric polarization due to the displacements of the ions within the unit AlGaAs unit cell. The resulting electric field has components $E_i$ that satisfy:[3]

$$D_i = \epsilon_0 \epsilon_s E_i + 2 \sum_k \hat{e}_{ik} \varepsilon_k \qquad \text{(S1)}$$

Here $\epsilon_0 = 8.85 \cdot 10^{-12}$ $C^2$ $N^{-1}$ $m^{-2}$ is the dielectric permittivity of vacuum and $\epsilon_s = 13.18 - 3.12x = 12.21$ (for $x$=0.31) is the dielectric constant of $Al_xGa_{1-x}As$ as a function of the Al concentration.[3] The piezoelectric coefficient of $Al_xGa_{1-x}As$ (for x=0.31) is $e_{14} = -0.16 - 0.07x$ C $m^{-2}$ $= -0.18$ C $m^{-2}$. $\varepsilon_{jk}$ is a rank 2 strain tensor that can be written in vector form as:[4]

$$\varepsilon_k = \begin{bmatrix} \varepsilon_{11} \\ \varepsilon_{22} \\ \varepsilon_{33} \\ \varepsilon_{23} \\ \varepsilon_{13} \\ \varepsilon_{12} \end{bmatrix}.$$

In the absence of external charges, the displacement field $D_i$ vanishes and Equation (S1) can then be solved to obtain the electric field.

$$E_i = -\frac{2 \sum_k \hat{e}_{ik} \varepsilon_k}{\epsilon_0 \epsilon_s} \qquad \text{(S2)}$$

In component form the electric field is given by:

$$E_i = -\frac{2}{\epsilon_0 \epsilon_s} \sum_{jk} \begin{bmatrix} 0 & 0 & 0 & e_{14} & 0 & 0 \\ 0 & 0 & 0 & 0 & e_{14} & 0 \\ 0 & 0 & 0 & 0 & 0 & e_{14} \end{bmatrix} \begin{bmatrix} \varepsilon_{11} \\ \varepsilon_{22} \\ \varepsilon_{33} \\ \varepsilon_{23} \\ \varepsilon_{13} \\ \varepsilon_{12} \end{bmatrix}. \qquad \text{(S3)}$$

The electric field along the depth direction of a [001] oriented thin film is $E_3 = -\frac{2 e_{14} \varepsilon_{12}}{\epsilon_0 \epsilon_s}$.

The magnitude of $E_3$, the 001 component of the electric field, due to gate depends on the arrangement of the gates and their orientation with respect to the crystallographic axes of the semiconductor. For a linear electrode oriented along the [010] direction, for example, the value of $\varepsilon_{12}$ is zero by symmetry and $E_3 = 0$.

The main text considers the case where the electrode is oriented along a [110] direction. In this case, symmetry considerations indicate that $\varepsilon_{12}$ is $\varepsilon_{12} = -\frac{1}{2}\varepsilon$, where $\varepsilon$ is the in-plane component of the strain induced by the electrode.

The potential difference between the gate and quantum dot due to piezoelectricity can be found by integrating the electric field from the surface to the depth of the 2DEG at a distance from the electrode corresponding to the location of the quantum dot. With an average in-plane strain of $4 \times 10^{-5}$, the magnitude of the potential difference due to piezoelectricity is $\Delta V_{PE} = 5$ mV.

## 2. Γ point deformation potential

The strain leads to a deformation potential that influences the effective Γ point Hamiltonian. In zinc-blende crystals, strain affects the electronic energy-band structure near the Γ point by shifting the minimum of the conduction band. This can be seen in the orbital-strain Hamiltonian given by:[3, 6]

$$\mathrm{H}_c^\Gamma = a_c^\Gamma \cdot \left(\varepsilon_{xx} + \varepsilon_{yy} + \varepsilon_{zz}\right).$$

Here $a_c^\Gamma = -11$ eV is the Bir-Pikus deformation potential of GaAs.[6] From the x-ray nanodifraction measurement, we estimate that the sum of the components of the strain at a distance from the electrode corresponding to the location of the quantum dot is $4 \times 10^{-6}$. The energy shift of the conduction band minimum at the quantum dot is $\Delta \mathrm{E}_c^\Gamma = -0.04$ meV.